%% file: eprint.tex
\documentclass[10pt]{article}
\usepackage{graphicx}

\def\LSP{\widetilde\chi^0_{1}}
\def\Ch{\widetilde\chi^{\pm}_{1}}
\def\sT{\widetilde{t}}
\def\sB{\widetilde{b}}
\def\sL{\widetilde{l}}
\def\santiT{\overline{\widetilde{t}}}
\def\santiB{\overline{\widetilde{b}}}
\def\deltam{\Delta m}

\def\Title#1{\begin{center} {\Large #1 } \end{center}}
\def\Author#1{\begin{center}{ \sc #1} \end{center}}
\def\Address#1{\begin{center}{ \it #1} \end{center}}

\newcommand\pubblock{\rightline{\begin{tabular}{l} Proceedings of the Fifth Annual LHCP\\ \pubnumber\\
         \pubdate  \end{tabular}}}

\newenvironment{Abstract}{\begin{quotation} \begin{center} 
             \large ABSTRACT \end{center}\bigskip 
      \begin{center}\begin{large}}{\end{large}\end{center} \end{quotation}}

\newenvironment{Presented}{\begin{quotation} \begin{center} 
             PRESENTED AT\end{center}\bigskip 
      \begin{center}\begin{large}}{\end{large}\end{center} \end{quotation}}

\def\Acknowledgements{\bigskip  \bigskip \begin{center} \begin{large}
             \bf ACKNOWLEDGEMENTS \end{large}\end{center}}

\input econfmacros.tex

\usepackage{color}

\newcommand\pubnumber{ CMS CR-2017/200 }

\newcommand\pubdate{\today}

\def\affiliation{
On behalf of the CMS Collaboration, \\
Universit\'e de Strasbourg, CNRS, IPHC UMR 7178, F-67000 Strasbourg, France}

\begin{document}

\large
\begin{titlepage}
\pubblock

\vfill
\Title{ Searches for third generation squarks with CMS }
\vfill

\Author{ Caroline Collard  }
\Address{\affiliation}
\vfill
\begin{Abstract}

Searches for the direct pair production of the third generation squarks are presented in this conference report, based on proton-proton collisions at a center-of-mass energy of 13 TeV and corresponding to an integrated luminosity of 35.9 fb$^{-1}$, collected by the CMS detector. New techniques have been developed to address different kinematical regimes. In absence of statistically significant deviations from the standard model background, limits have been derived in terms of simplified model spectra, excluding for example a stop up to a mass of 1.05~TeV when it decays to a top and a massless neutralino. 

\end{Abstract}
\vfill

\begin{Presented}
The Fifth Annual Conference\\
 on Large Hadron Collider Physics \\
Shanghai Jiao Tong University, Shanghai, China\\ 
May 15-20, 2017
\end{Presented}
\vfill
\end{titlepage}
\def\thefootnote{\fnsymbol{footnote}}
\setcounter{footnote}{0}
%

\normalsize 


\section{Introduction}

The motivations to search for the third generation squarks, namely the stop $\sT$ and sbottom $\sB$, in supersymmetry (SUSY) are based on two observations.
The first one comes from the recent Higgs boson discovery at a mass of 125 GeV~\cite{higgs1,higgs2}. Natural SUSY scenarios can protect the mass of the Higgs boson by cancelling 
the divergence of quantum loop corrections thanks to the contributions involving supersymmetric particles, and in that case the stop and the sbottom are expected to be light. The second important 
observation is the presence of dark matter in the universe. Working under R-parity conservation (RPC) allows the lightest supersymmetric particle (LSP) to be stable and it therefore could provide a candidate for the dark matter.  In the following, we only consider the case of the neutralino ($\LSP$) to be the LSP.  Another consequence of RPC is that sparticles are pair produced. This report focusses on results targeting the direct production of $\sT \santiT$  and $\sB \santiB$ using the data collected by the CMS detector~\cite{cms}, and corresponding to an integrated luminosity of 35.9~fb$^{-1}$. Different decay chains of stop or bottom up to the LSP are considered. The decay products depend on the masses of the involved sparticles, i.e. m($\sT$ or $\sB$),  m($\LSP$), and other sparticles available in between in the mass spectrum, leading to different final states with different kinematics. In all the presented analyses, Simplified Model Spectra (SMS) are used to guide our studies and interpret the results.

\section{Techniques and analysis strategies in SUSY}

To address the different kinematical regimes, a wide range of techniques are used. Some of them are very standard in SUSY like the use of different physics objects (like leptons, jets, b jets, 
missing transverse energy (MET)), the reconstruction of various masses or kinematical variables, etc. \\

\begin{figure}[htb]
\centering
\includegraphics[height=1.8in]{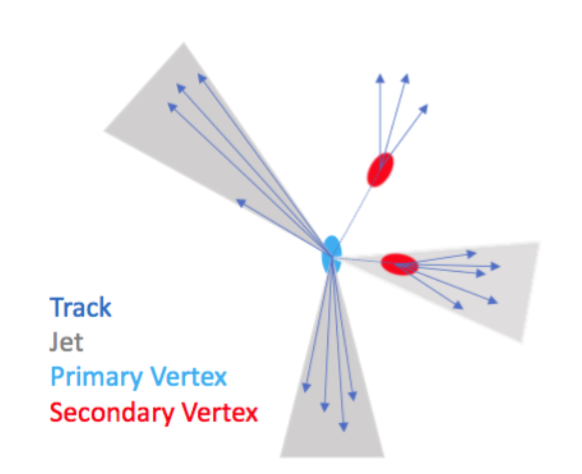}
\includegraphics[height=2.0in]{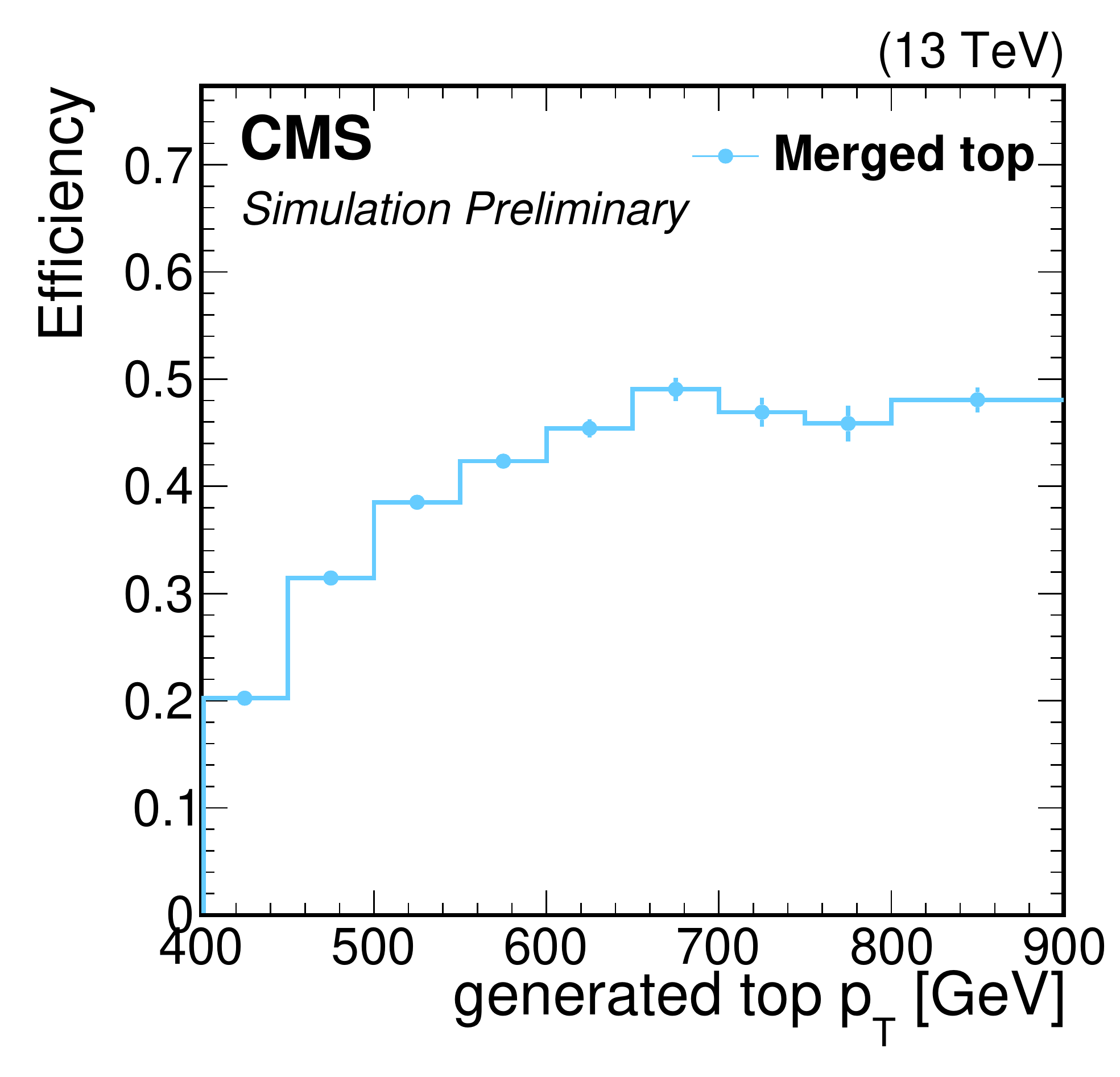}
\includegraphics[height=2.0in]{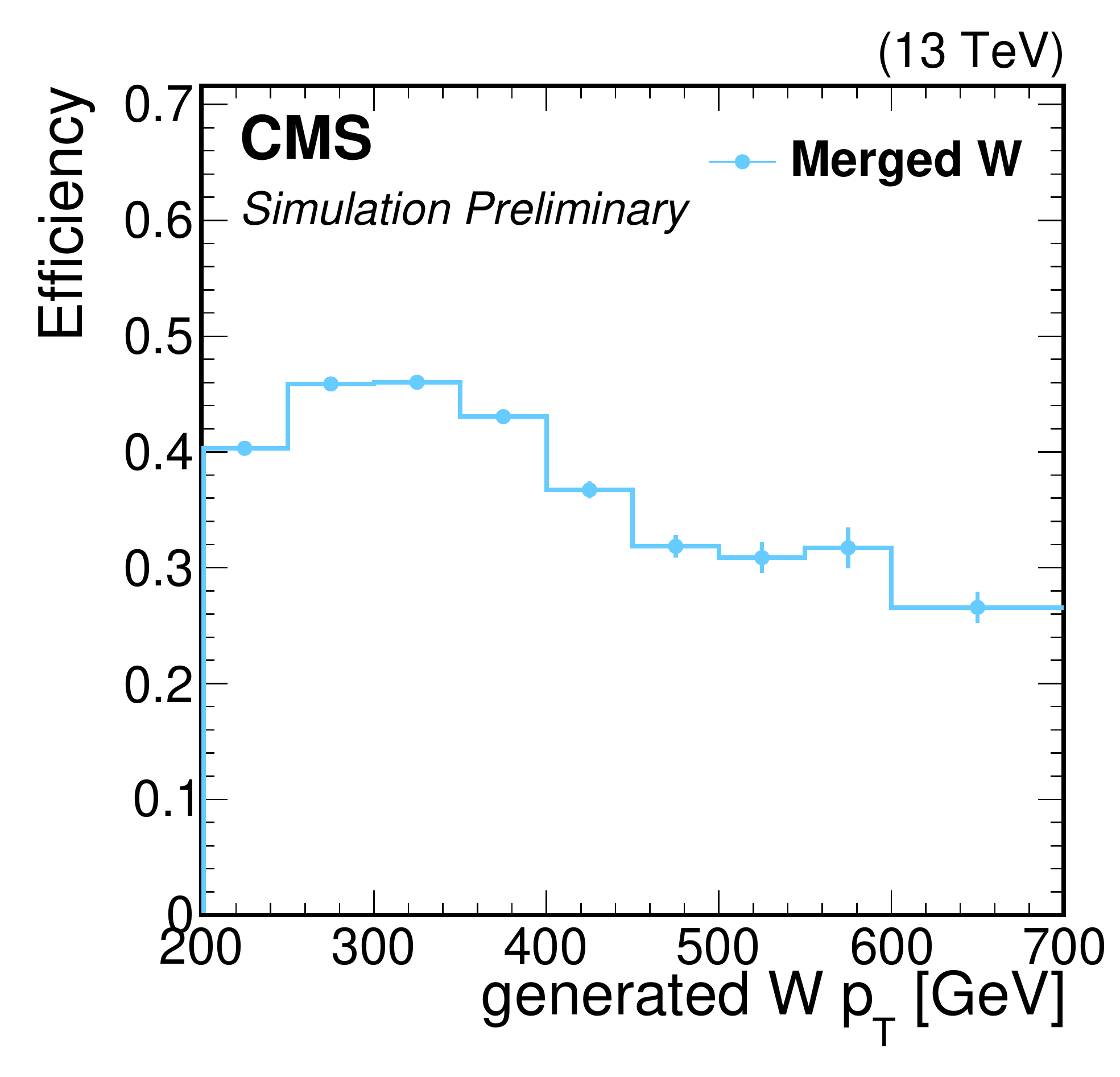}
\caption{Sketch of soft b-tagging (left). Efficiency to correctly identify a merged top (middle) or W  boson (right)  as a function of the  $p_T$ of the generated top quark or W boson~\cite{sus16049}.}
\label{fig:figure1}
\end{figure}

In compressed regime characterised by a low $\deltam$ value (with $\deltam$ =m($\sT$ or $\sB$) - m($\LSP$)), the decay products may be rather soft and the researched signal could look as standard model
background. In order to improve the sensitivity of the searches, new techniques have been put in place in some of the analyses. A soft b-tagging tool~\cite{sus16049} has been designed, based on the 
reconstruction of a secondary vertex, not associated to a jet, as shown on Figure~\ref{fig:figure1} (left). This allows to identify b jets with $p_T$ values lower than 20 or 25 GeV depending on the analyses. 
A c-tagger has also been developed to identify c jets while rejecting b and light jets~\cite{btv16001}. The presence of additional jets coming from initial state radiations (ISR) is often required at low $\deltam$ value to boost final state system and provide enough MET in the event. Lastly, leptons labeled as soft because of their low $p_T$ value (3.5 $<p_T <$ 30 GeV) bring also interesting information, and dedicated triggers have been put in place to ensure the storage of these particular events~\cite{sus16048}. \\

For very large values of $m(\sT)$, the decay products are boosted and dedicated W and top taggers have been designed~\cite{sus16049,sus16050}. The performance of some of them are presented on Figure~\ref{fig:figure1} (middle and right). Lowering the jet multiplicity requirement may also be helpful to address the boosted areas~\cite{sus16051}.\\

Based on these techniques, the categorisation of events is performed through the design of various Signal Regions (SR), which are optimised for specific signals to target defined regions of the phase space and to disentangle signals and backgrounds. Then the different background contributions are mostly estimated with data-driven techniques, based on data yields measured in Control Regions (CR) orthogonal to the SR (for example by requesting one additional lepton or inverting the b-tagging selection), and then multiplied by transfer functions computed with Monte Carlo simulations (MC) to pass from CR to SR. Because of lack of statistics for the rare backgrounds, their yields in SR are directly taken from MC. Finally the results are interpreted. If there is no statistically significant deviation observed in data with respect to the background prediction, exclusion limits are extracted. 

\section{Searches for stop and sbottom}

In this report, we concentrate on three different stop searches, depending on the lepton multiplicity in the final state: the fully hadronic (0$\ell$)~\cite{sus16049}, the semileptonic (1$\ell$)~\cite{sus16051}  and the dileptonic  (2$\ell$)~\cite{sus17001} analyses.  \\

First, the 0$\ell$ analysis~\cite{sus16049} is based on a preselection requesting 0 lepton, at least 2 jets and MET$>$ 250 GeV. The backgrounds are composed by t$\bar{\rm{t}}$, W+jets and single top events in which one generated lepton is lost, by events with a Z boson decaying invisibly, and by small contributions from QCD multijet production and rare processes. The categorisation addresses separately the low and high $\Delta m$ regimes, depending on the transverse mass $M_T (b_{1,2},\rm{MET})$  in the event. In total, 104 SR are used based on cuts on multiplicities of jets, b jets and soft b jets, on the transverse momenta of the ISR system and of the b jets, and finally on MET.\\

\begin{figure}[htb!]
\centering
\includegraphics[height=1.85in]{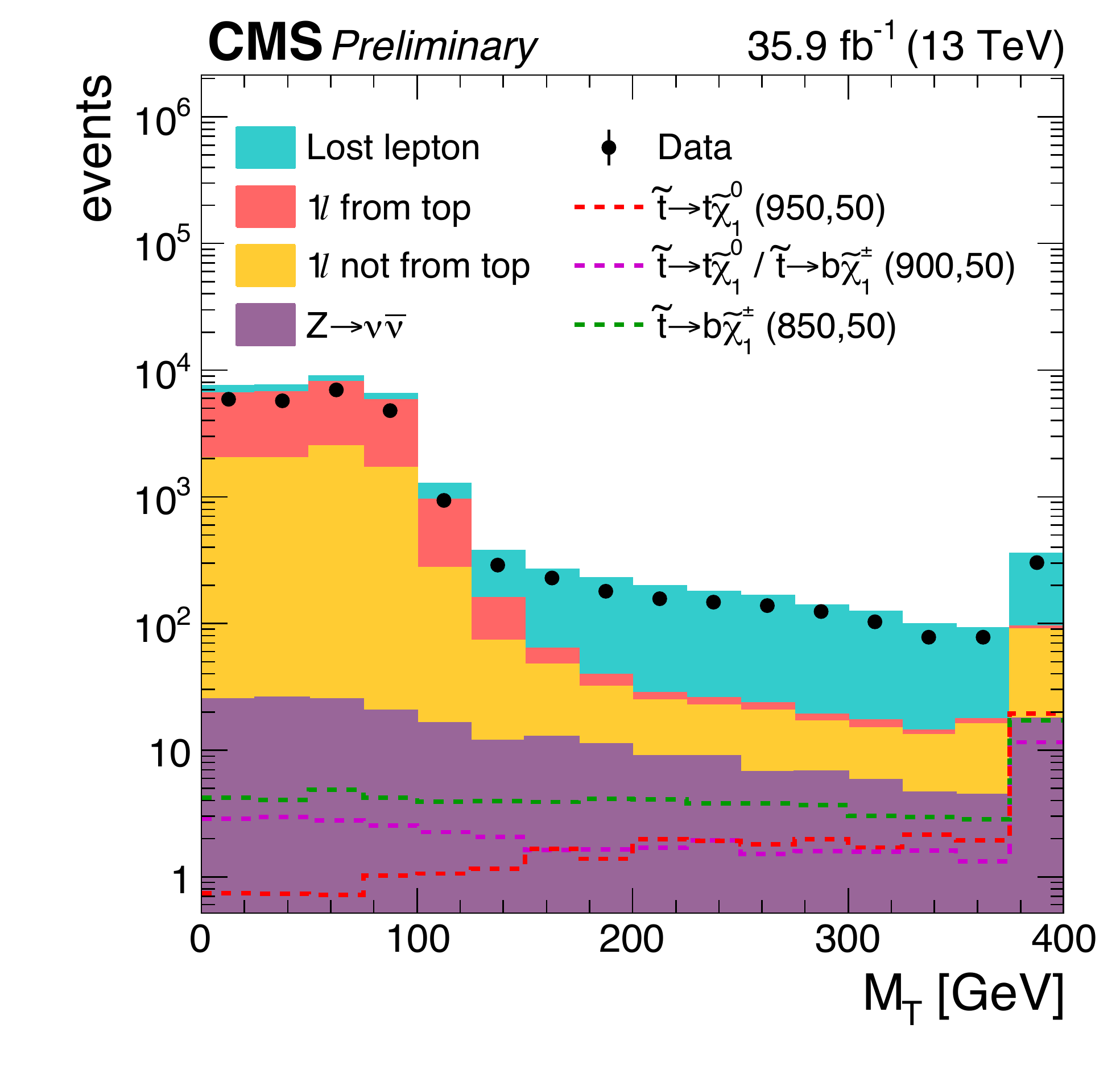}
\includegraphics[height=1.85in]{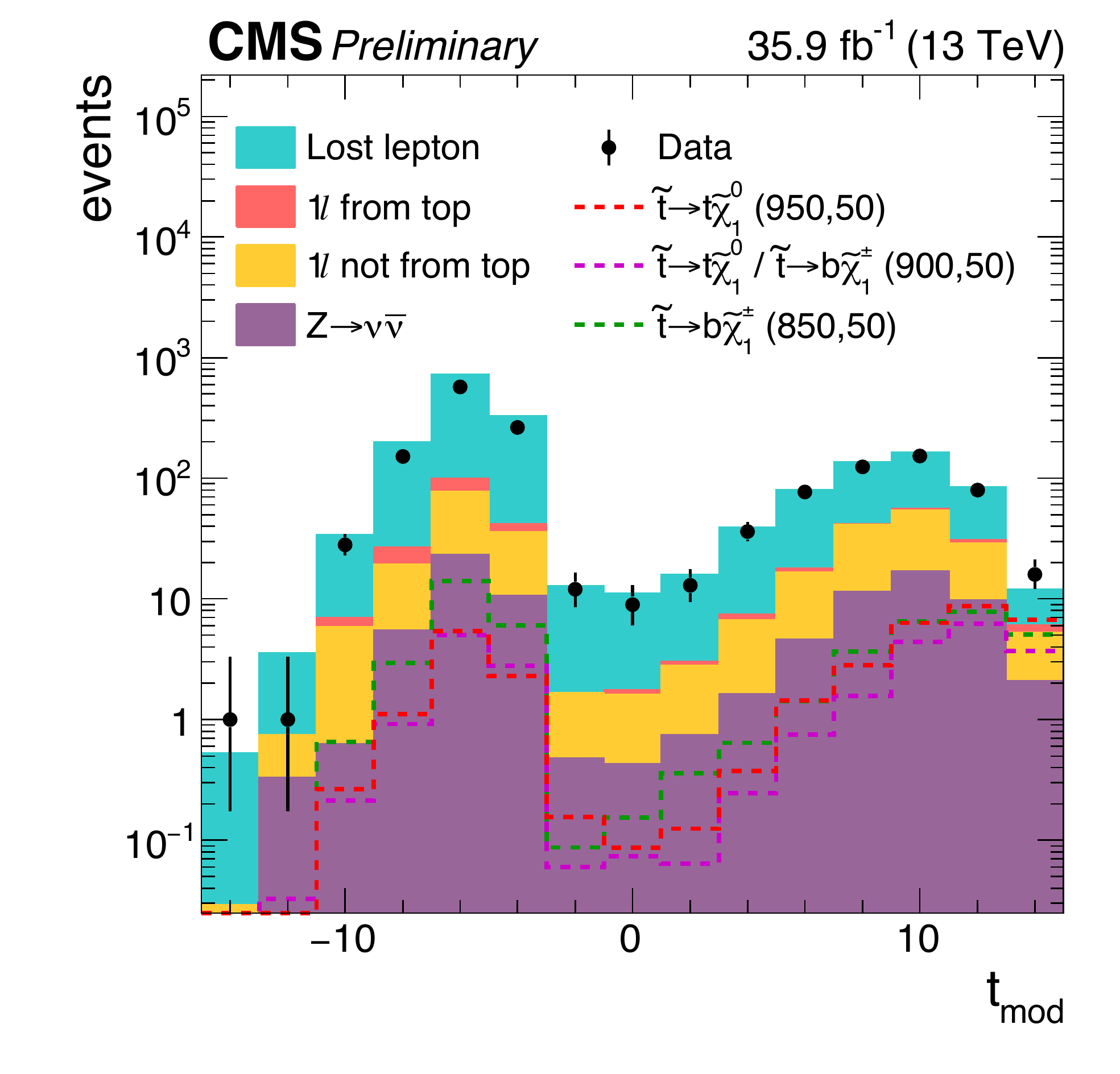}\\
\includegraphics[height=2.0in]{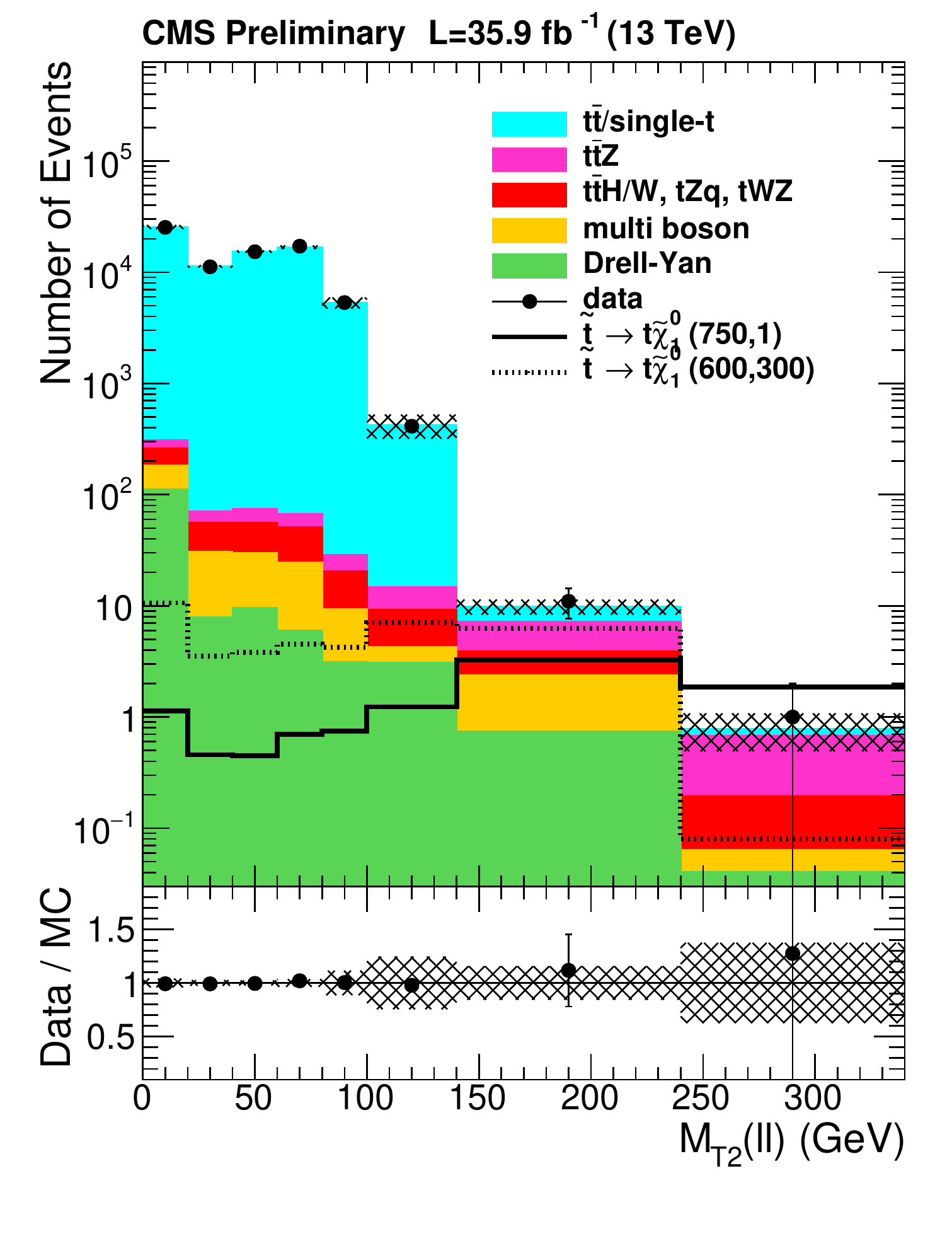}
\includegraphics[height=2.0in]{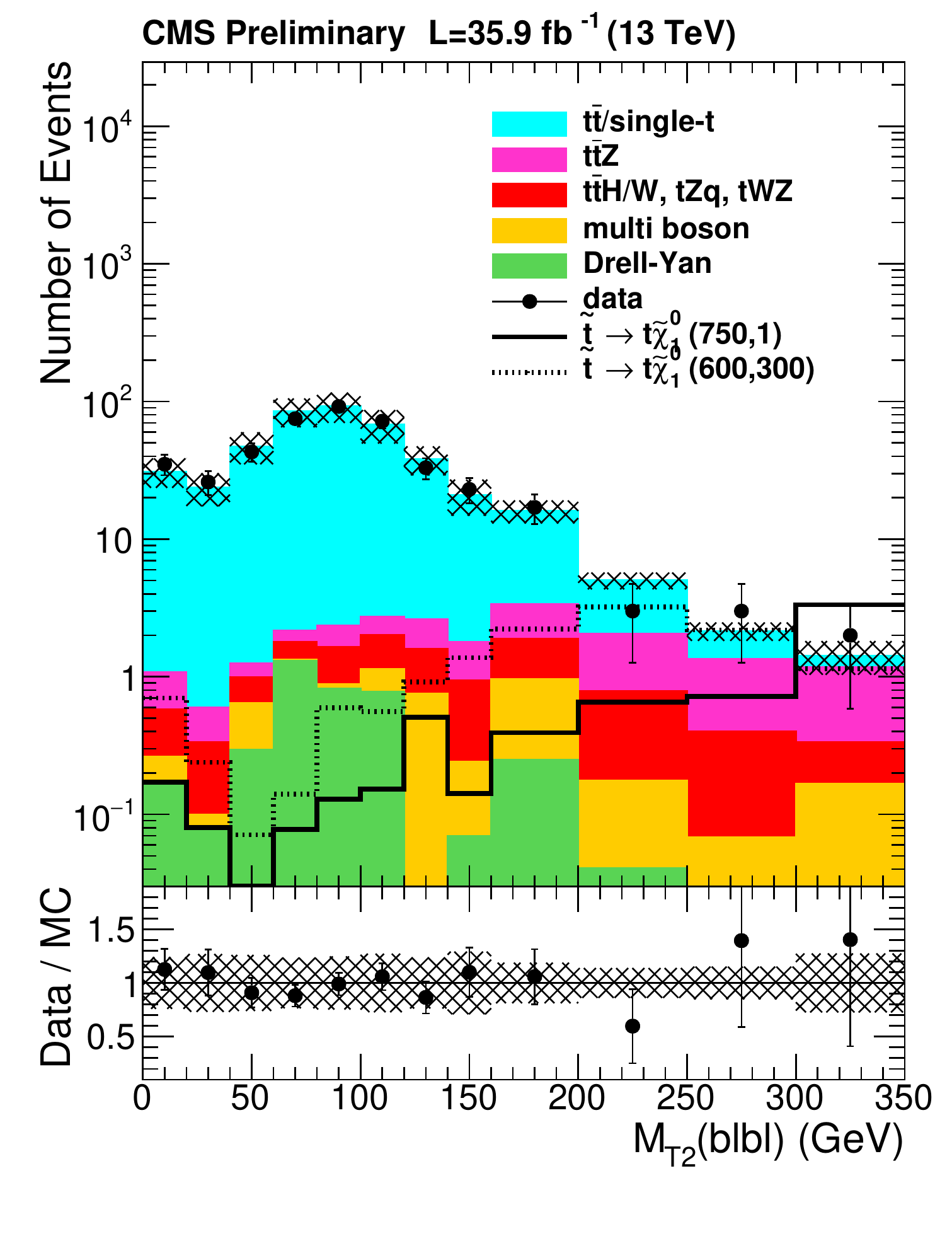}
\caption{(top) Distribution of $M_T$ (left) and the modified topness variable (right) for events passing the preselection of the 1$\ell$ analysis~\cite{sus16051}. (bottom) Distribution of $M_{T2}(\ell \ell)$ (left) and $M_{T2}(b\ell b\ell)$ (right) for events passing the preselection of the 2$\ell$ analysis~\cite{sus17001}.}
\label{fig:figure2}
\end{figure}

Secondly, the 1$\ell$ analysis~\cite{sus16051} requests exactly one electron or muon, at least 2 jets with at least 1 b jet among them, MET$>$ 250 GeV and $M_T>$150 GeV,
the discriminating power of the $M_T$ variable being presented on Figure~\ref{fig:figure2} (top left). Here also the main backgrounds come from processes with a lost lepton (when two are generated) or with an invisible Z boson, 
in addition to some more rare backgrounds. The interesting variables used for the categorisation are the modified topness (see Figure~\ref{fig:figure2} (top right)), $M(\ell,b)$, 
the jet multiplicity and MET. The compressed region is addressed with a dedicated ISR selection. In total, 31 SR have been designed. \\

Lastly, the 2$\ell$ analysis~\cite{sus17001} with exactly 2 leptons (electrons or muons) outside the Z peak imposes at least 2 jets with at least 1 b jet among them, 
$M_{T2}(\ell \ell)>$ 100 GeV, MET $>$ 80 GeV and MET$/\sqrt{H_T}>5 \rm{GeV}^{-1}$. Processes with at least one top quark (t$\bar{\rm{t}}$, single top, t$\bar{\rm{t}}$Z, t$\bar{\rm{t}}$H, t$\bar{\rm{t}}$W, tZq, tWZ) contribute to the 
backgrounds, as well as multiboson and Drell Yan events. Figure~\ref{fig:figure2} (bottom) shows two of the important discriminating variables used in the categorisation, 
namely $M_{T2}(\ell \ell)$ and $M_{T2}(b\ell b\ell)$, the other selections being based on the lepton flavours and on MET. In total, 26 SR are used.\\

\begin{figure}[htb!]
\centering
\includegraphics[height=2.23in]{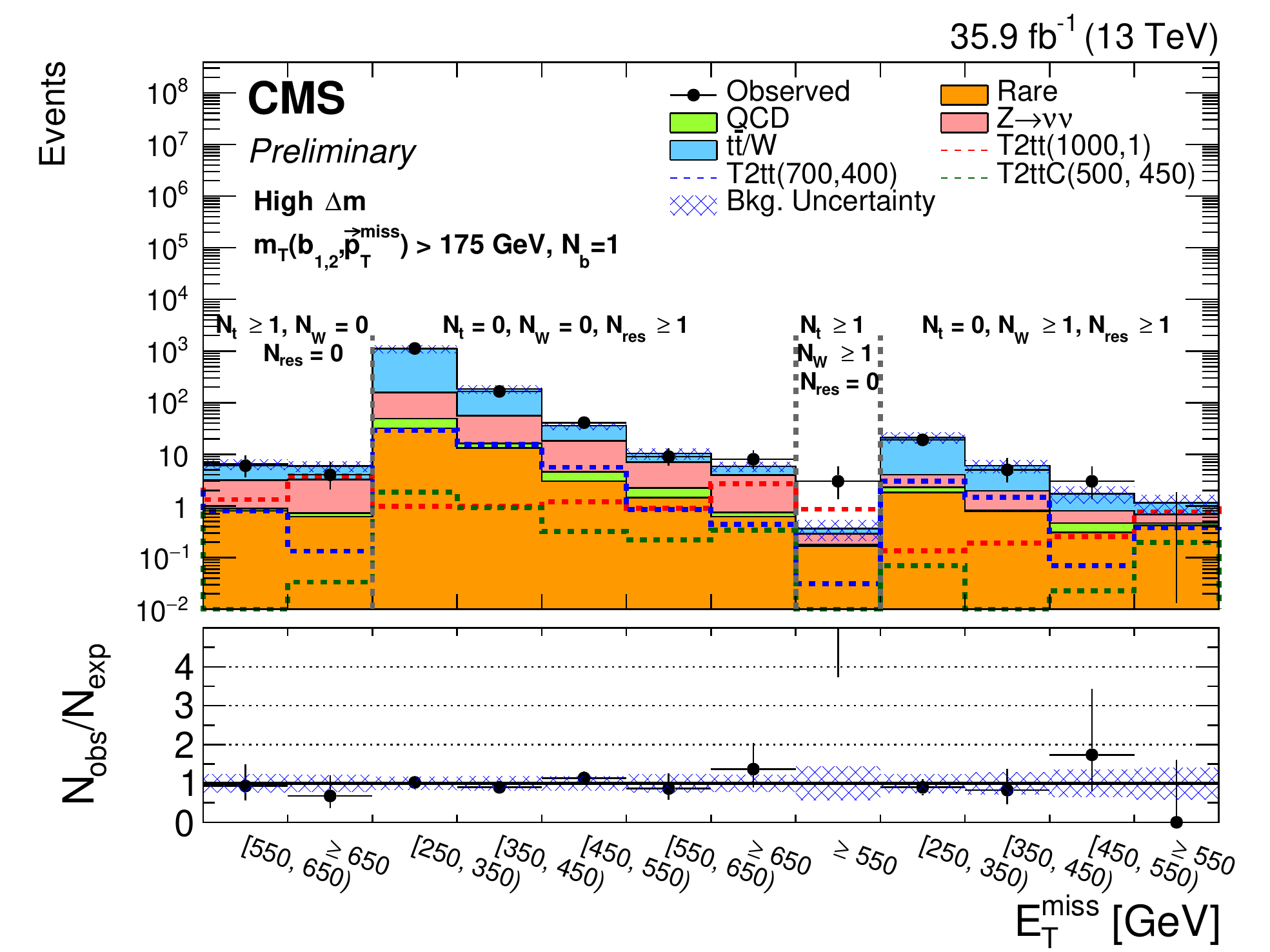}
\includegraphics[height=2.23in]{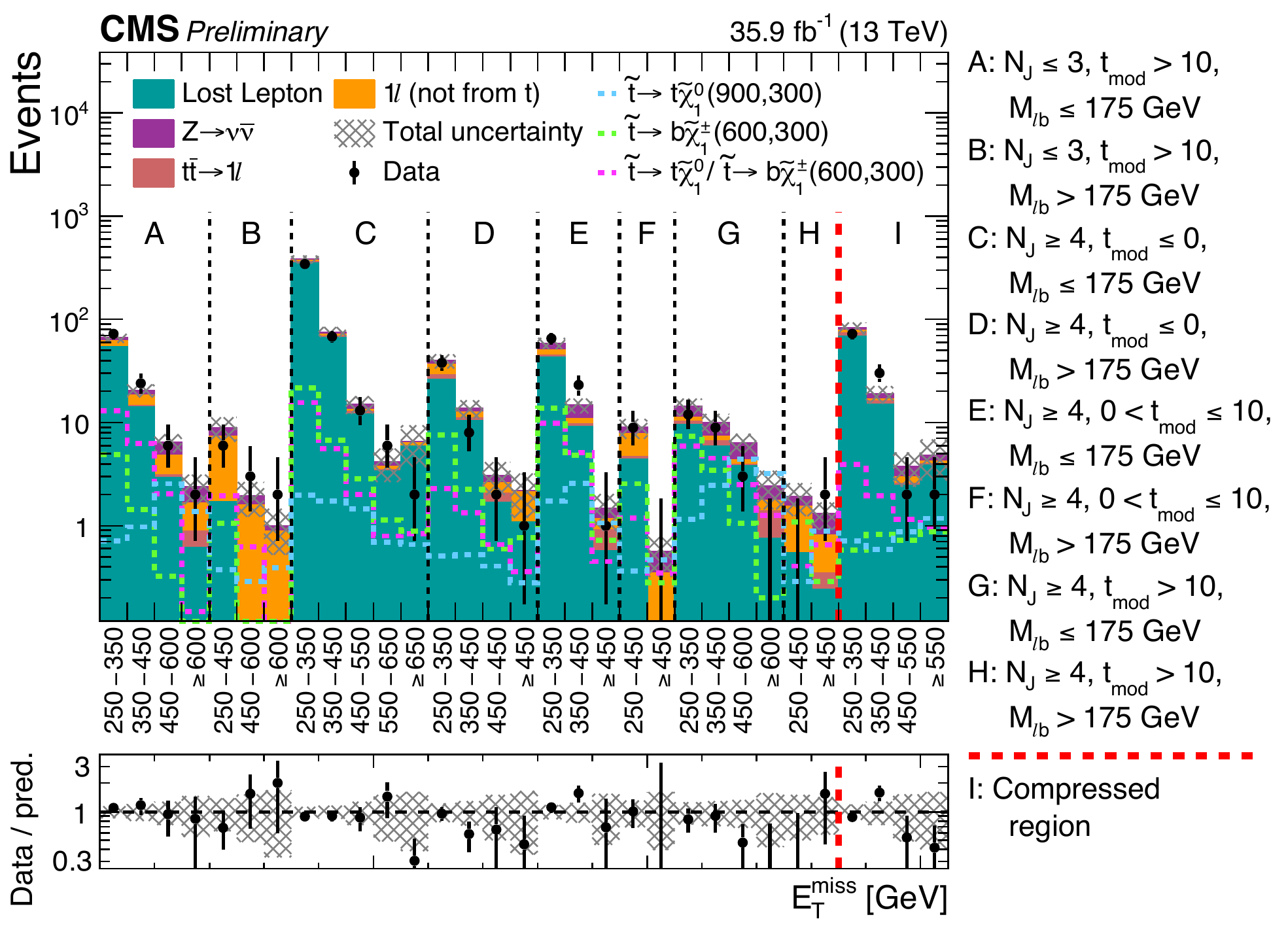}
\caption{Observed data events and SM background predictions in a subset of signal regions of the 0$\ell$ analysis~\cite{sus16049} (left) and in all the signal regions of the 1$\ell$ analysis~\cite{sus16051} (right).}
\label{fig:figure3}
\end{figure}

Figure~\ref{fig:figure3} shows some comparisons of data and background yields for the 0$\ell$ and 1$\ell$ stop analyses. Note that the excess observed in the 8$^{th}$ bin in left figure is of 1.9 $\sigma$, 
and apart from another excess of 2.3 $\sigma$ in a low $\Delta m$ SR of the 0$\ell$ analysis, no other deviation has been observed. \\

\begin{figure}[htb!]
\centering
\includegraphics[height=2.23in]{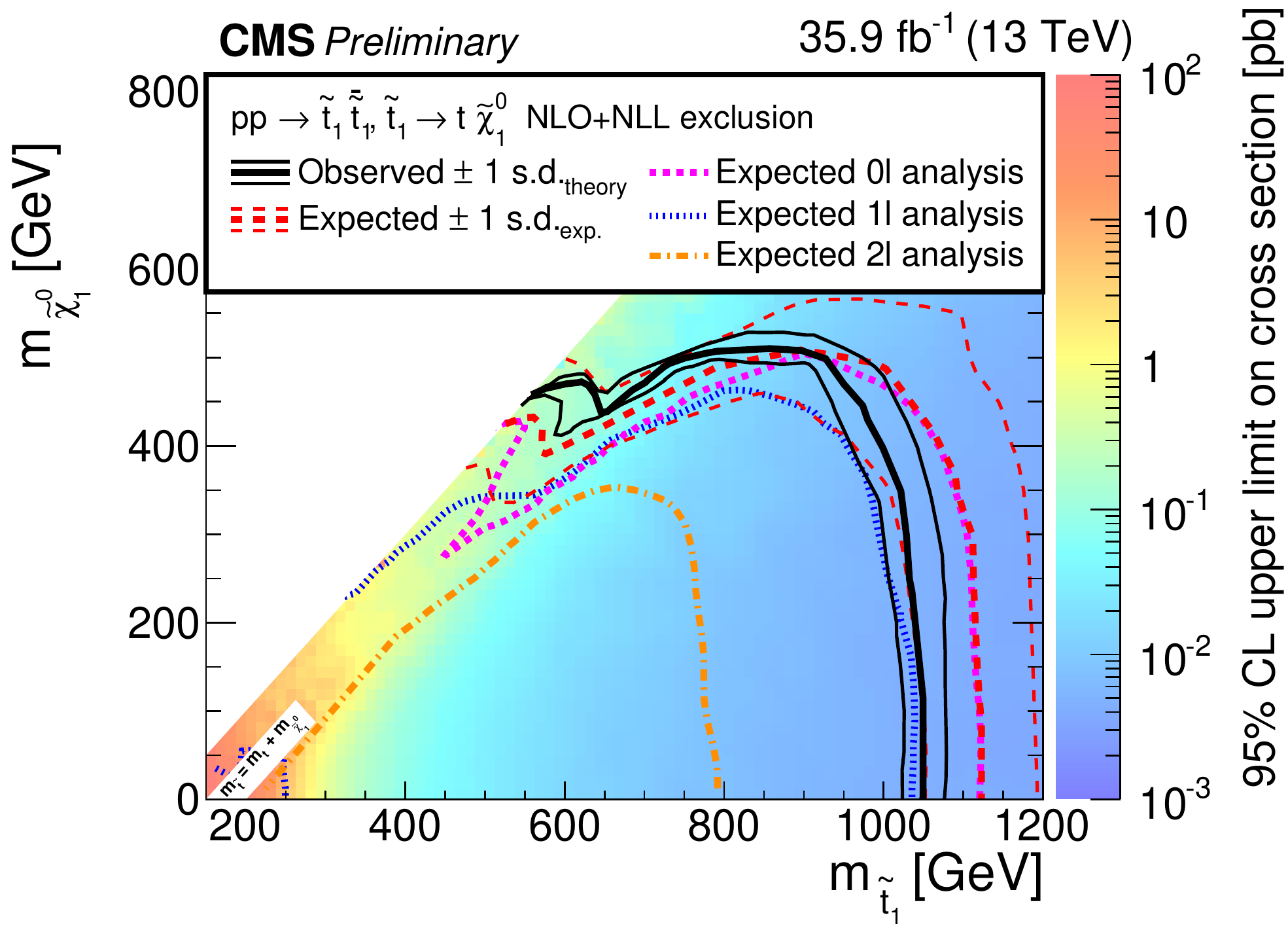}
\includegraphics[height=2.23in]{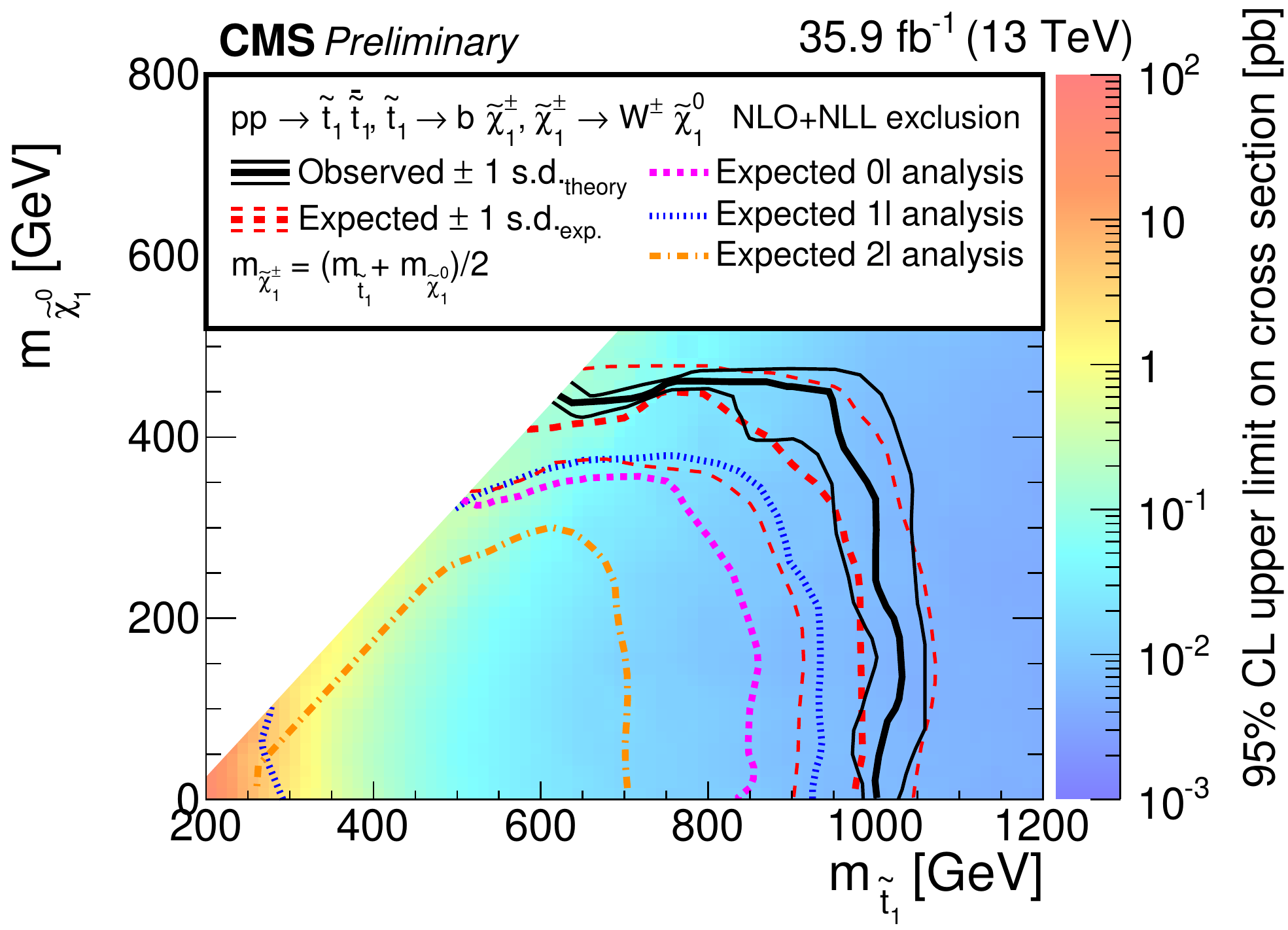} \\
\includegraphics[height=2.23in]{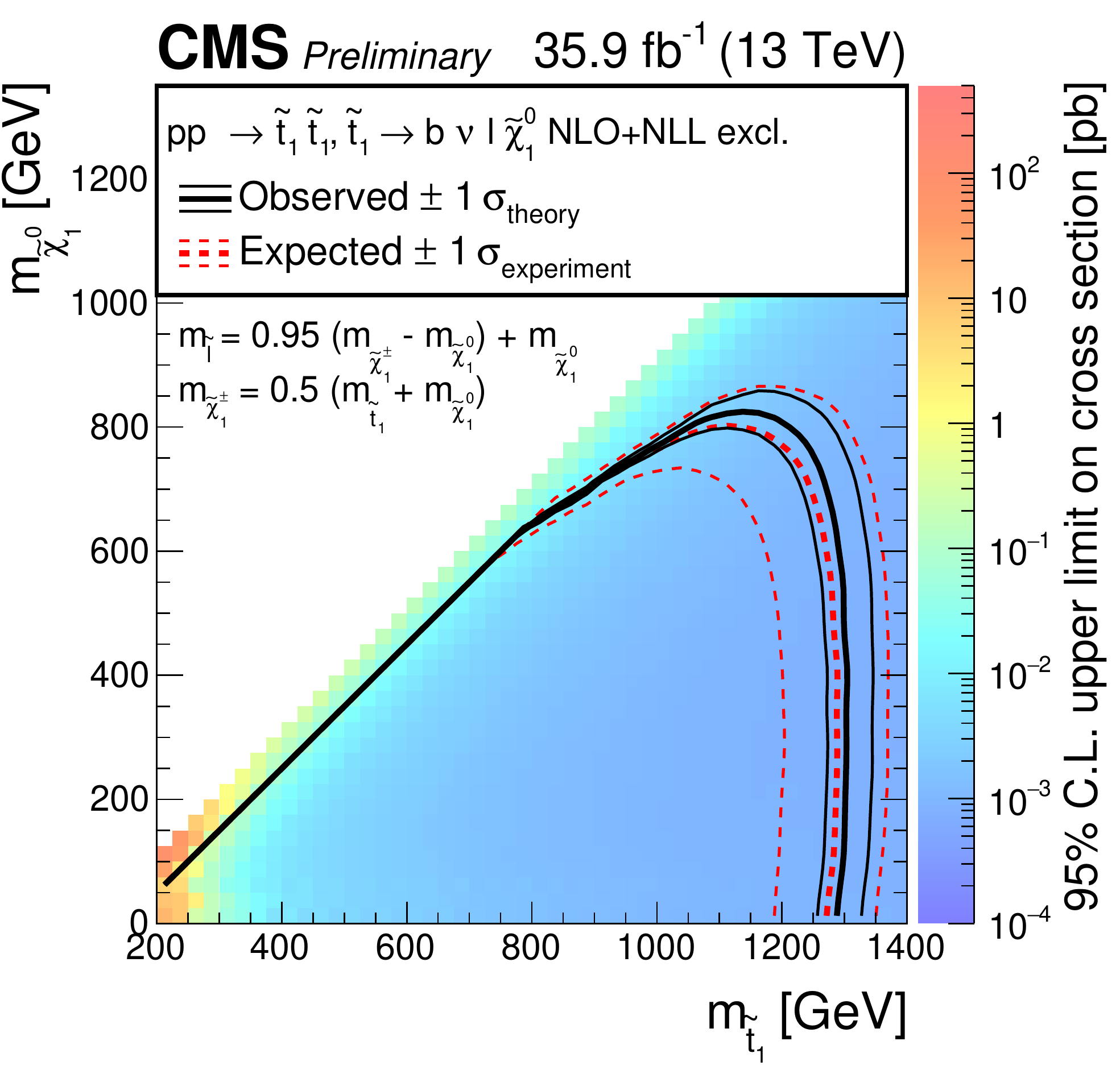}
\includegraphics[height=2.23in]{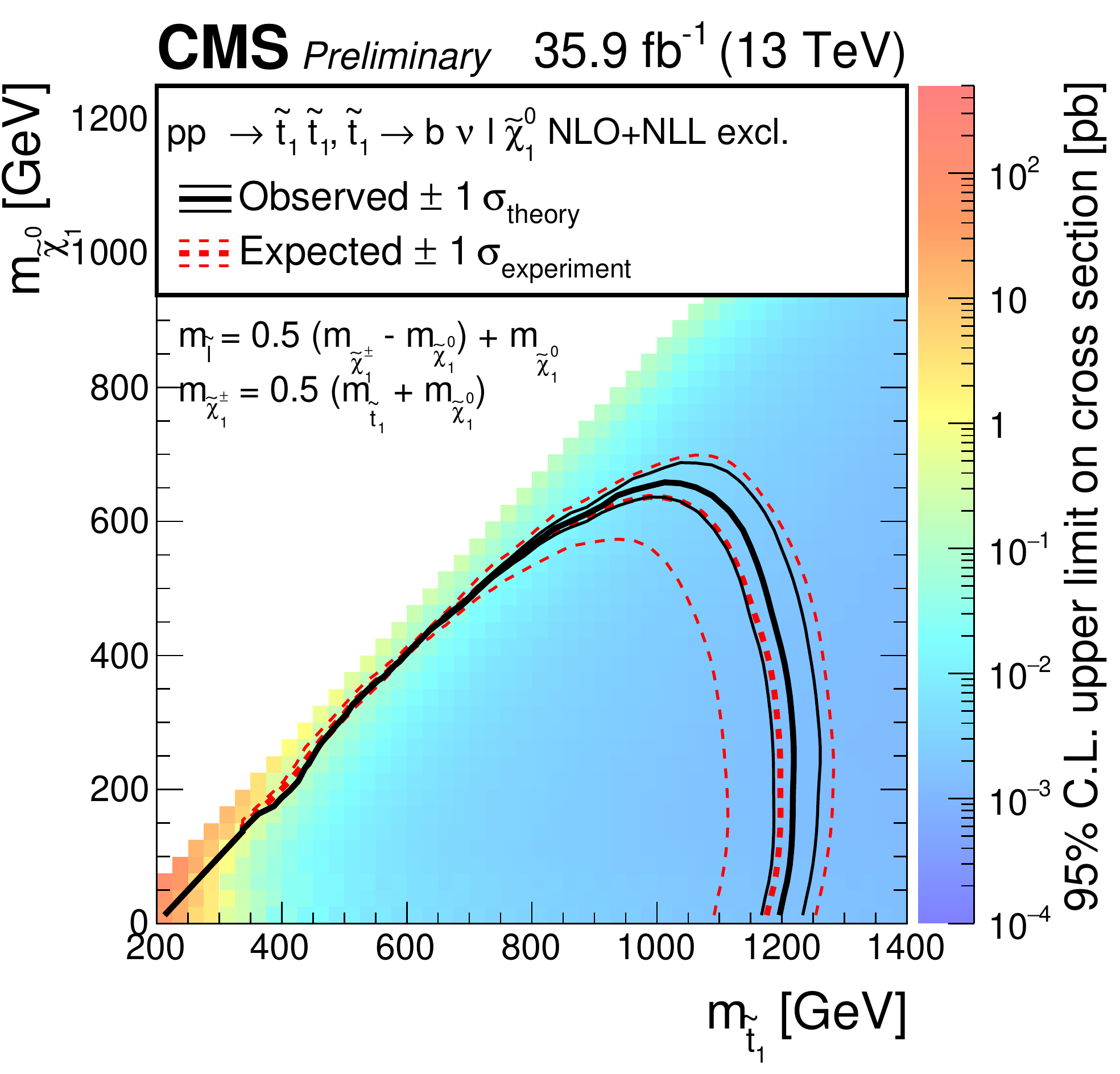}\\
\includegraphics[height=2.23in]{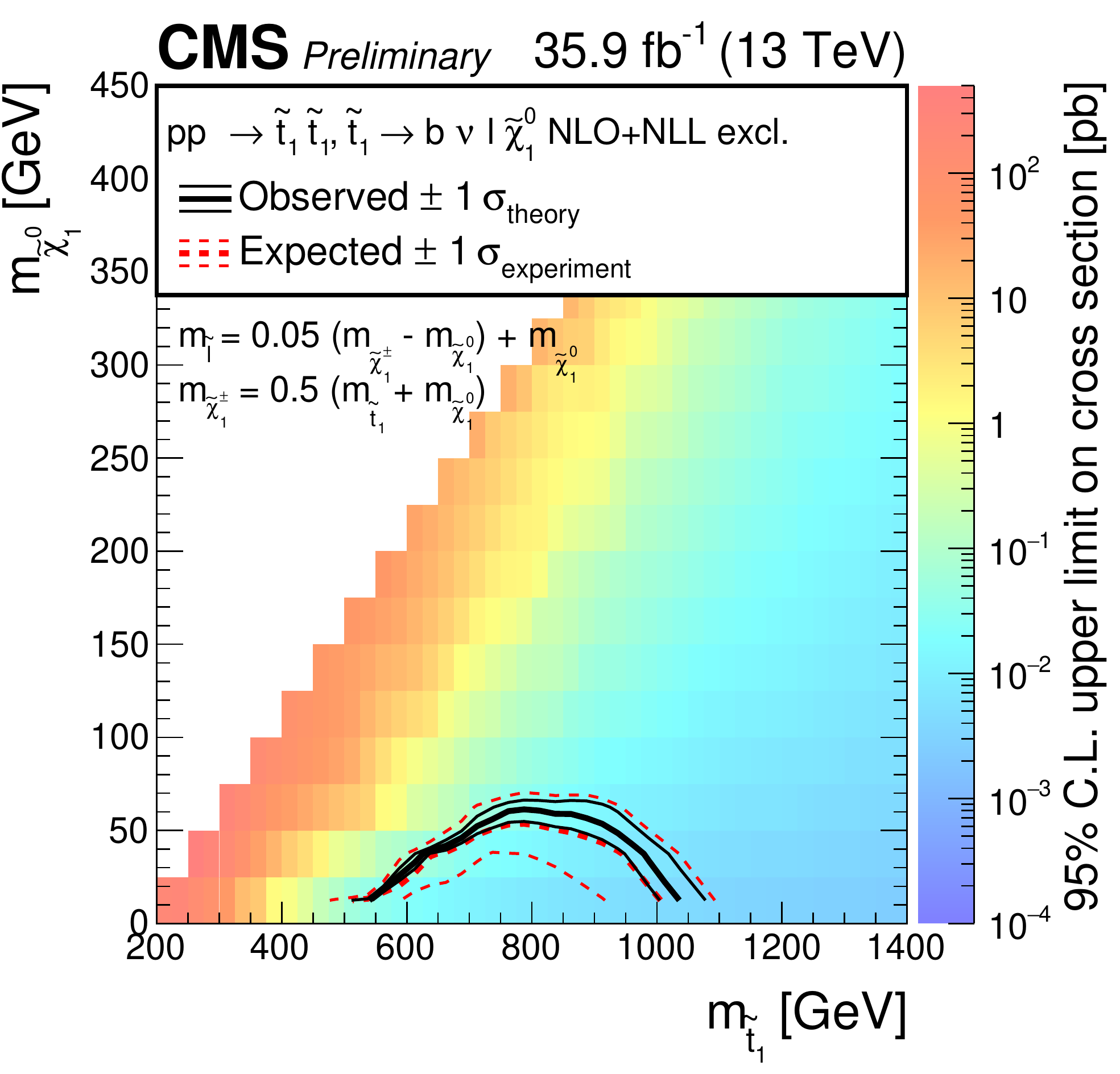}
\includegraphics[height=2.23in]{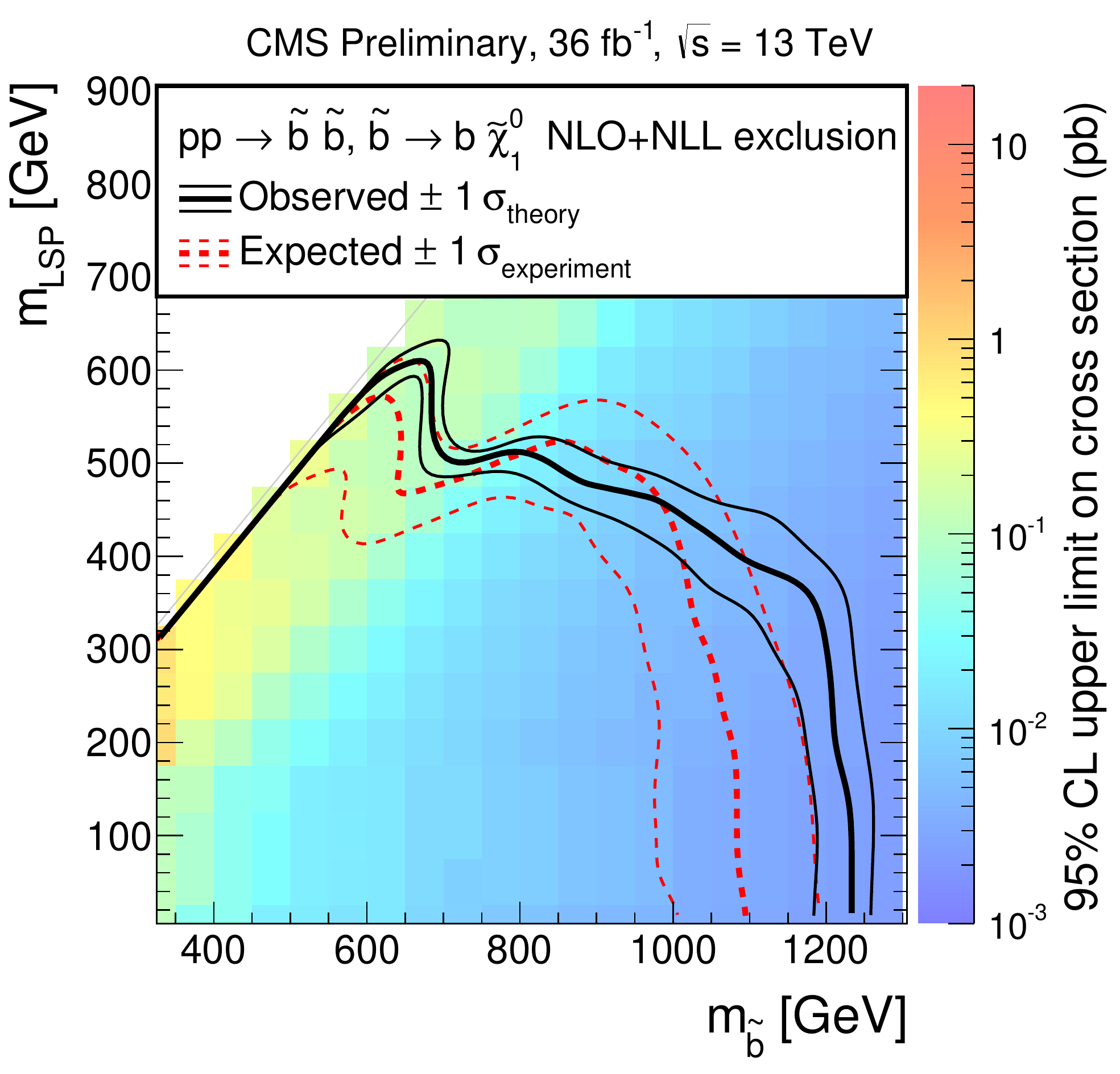}
\caption{Exclusion limits at 95\% CL for different SMS models. (top) Direct stop pair production with the decay mode  $\sT \rightarrow t + \LSP$ (left) and $\sT \rightarrow b + \Ch$ with $\Ch \rightarrow W + \LSP$ and $m(\Ch) = 0.5*(m(\sT) + m(\LSP))$ (right). In addition to the curves for the combination~\cite{combination}, the expected limits from the three individual analyses are also displayed in magenta short-dashed, blue dotted, and long-short-dashed orange for the 0$\ell$~\cite{sus16049}, 1$\ell$~\cite{sus16051} and 2$\ell$~\cite{sus17001} analyses, respectively. (middle line and bottom left) Direct stop pair production with the cascade decay mode $\sT \rightarrow b + \Ch  \rightarrow \nu + \sL  \rightarrow \ell + \LSP$ for three different m($\sL$) scenarios~\cite{sus17001}. (bottom right) Direct sbottom  pair production with the decay mode  $\sB \rightarrow b + \LSP$~\cite{sus16032}. In all plots, the color indicates the 95\% CL upper limit on the cross section times branching fraction at each point in the m($\sT$ or $\sB$)-m($\LSP$) plane. The area to the left of and below the thick black curve represents the observed excluded region at 95\% CL, while the dashed red lines indicate the expected limits at 95\% CL and their $\pm$1 experiment standard deviation uncertainties. The thin black lines show the effect of the theoretical uncertainties theory on the signal cross section.}
\label{fig:figure4}
\end{figure}

The results have been interpreted for different decay modes of the stop, the exclusion limits are presented in Figure~\ref{fig:figure4}. 
In the case of $\sT \rightarrow t + \LSP$ (top left), most of the sensitivity comes from the 0$\ell$ analysis at high $m(\sT)$. 
For a massless $\LSP$, $\sT$~masses are excluded up to 1.05 TeV.  The combination of the 3 analyses~\cite{combination} really improves the sensitivity at low $\Delta m$. 
For $\sT \rightarrow b + \Ch$ (top right) with $\Ch \rightarrow W + \LSP$ and m($\Ch$) = $0.5*$(m($\sT$) + m($\LSP$)), 
most of the sensitivity comes this time from the 1$\ell$ analysis. The combined limit allows to exclude a $\sT$  mass of 1~TeV for m($\LSP$)$\sim$0 GeV.
In the case of the cascade decay mode $\sT \rightarrow b + \Ch  \rightarrow \nu + \sL  \rightarrow \ell + \LSP$ (middle and bottom left), the exclusion area is quite sensitive to the slepton ($\sL$) mass. \\

The search for sbottom is performed in a purely hadronic final state~\cite{sus16032}. The limits displayed on Figure~\ref{fig:figure4} (bottom right) show an exclusion of the $\sB$ mass up to 1.225 TeV for $m(\LSP)<$ 100 GeV. 

\section{Conclusions}

A wide variety of searches for the third generation squarks is performed in CMS. No sign of SUSY arose so far and limits have been set in terms of SMS.
An important improvement in sensibility is observed in the region at high m($\sT$) with a $\sim$150 GeV gain with respect to the results presented at the ICHEP 2016 conference~\cite{ichep2016} showing the importance of boosted object tagging.
The compressed spectra have benefitted from the design of dedicated searches using new techniques like for example soft b-tagging or c-tagging.

\Acknowledgements
I would like to acknowledge the organizers for this very pleasant and interesting conference. I am grateful for all the lively discussions I had in Shanghai.

\end{document}

%% file: econfmacros.tex
\def\beq{\begin{equation}}
\def\eeq#1{\label{#1}\end{equation}}
\def\eeqn{\end{equation}}

\def\beqa{\begin{eqnarray}}
\def\eeqa#1{\label{#1}\end{eqnarray}}
\def\eeqan{\end{eqnarray}}

\let\bar=\overbar

\def\Dslash{\not{\hbox{\kern-4pt $D$}}}
\def\dslash{\not{\hbox{\kern-2pt $\del$}}}

\def\msb{{\bar{\ssstyle M \kern -1pt S}}}

